# High-Throughput Large-Area Automated Identification and Quality Control of Graphene and Few-Layer Graphene Films

Craig M. Nolen,[1] Giovanni Denina,[2] Desalegne Teweldebrhan,[1] Bir Bhanu,[2] and Alexander A. Balandin[1,*]

[1]Nano-Device Laboratory, Department of Electrical Engineering and Materials Science and Engineering Program, Bourns College of Engineering, University of California – Riverside, Riverside, California 92521, USA

[2]Visualization and Intelligent Systems Laboratory, Department of Electrical Engineering, Bourns College of Engineering, University of California – Riverside, Riverside, California 92521, USA

[*]Corresponding author; E-mail address: balandin@ee.ucr.edu ; web-site: http://ndl.ee.ucr.edu/

**Abstract**

Practical applications of graphene require a reliable high-throughput method of graphene identification and quality control, which can be used for large-scale substrates and wafers. We have proposed and experimentally tested a fast and fully automated approach for determining the number of atomic planes in graphene samples. The procedure allows for *in situ* identification of the borders of the regions with the same number of atomic planes. It is based on an original image processing algorithm, which utilizes micro-Raman calibration, light background subtraction, lighting non-uniformity correction, and the color and grayscale image processing for each pixel. The outcome of the developed procedure is a pseudo-color map, which marks the single-layer and few-layer graphene regions on the substrate of any size that can be captured by an optical microscope. Our approach works for various substrates, and can be applied to the mechanically exfoliated, chemically derived, deposited or epitaxial graphene on an industrial scale.

C. M. Nolen, G. Denina, D. Teweldebrhan, B. Bhanu and A.A. Balandin, University of California – Riverside, 2010

Graphene is a two-dimensional (2-D) crystal of $sp^2$-bonded carbon atoms (1). Mechanical exfoliation of graphene (1) led to the discovery of graphene's extraordinary electronic (1), thermal (2), optical (3-5) and mechanical properties (6). Strongly dependent on the number of atomic planes, physical characteristics of few-layer graphene (FLG) are different from those of single layer graphene (SLG). For example, SLG reveals electron mobility in the range from ~40,000 to 400,000 $cm^2V^{-1}s^{-1}$ (7) and intrinsic thermal conductivity above ~3000 W/mK for large suspended flakes (2, 8-11) while bilayer graphene (BLG) exhibits electron mobility in the range from ~3000 to 8000 $cm^2V^{-1}s^{-1}$ (12) and intrinsic thermal conductivity near ~2500 W/mK (13, 14). The electronic, thermal and optical properties of FLG approach those of bulk graphite as the number of atomic layers exceeds approximately ten layers (15, 16). The optical transparency of FLG is also a strong function of the number of layers (3-5). The one-atom thickness of graphene and its optical transparency (only ~2.3% absorption per layer (3)) make graphene identification and counting the number of atomic planes in FLG extremely challenging.

Recent progress in the chemical vapor deposition (CVD) growth of graphene led to fabrication of large-area graphene layers that are transferable onto various insulating substrates (17, 18). The CVD grown graphene layers of up to 30 inches in size on cheap flexible substrates have been demonstrated (17). Various other methods of graphene synthesis were reported (19, 20, 21). It is reasonable to expect, in the near future, the emergence of graphene growth techniques on insulating substrates, which would allow one to avoid the graphene transfer steps. The fusion of the large-area graphene on cheap, transparent, flexible substrates with graphene-based OLED technology is expected to lead to major practical applications (22). However, as larger area graphene becomes available, quality control remains as an important factor limiting further progress in graphene research and applications. For all these reasons, it is important to develop a fast scalable method for determining the number of atomic planes in synthesized graphene or FLG. The crucial feature of this method, which would allow for industry applications, should be its suitability for large-area substrates (i.e., lateral dimensions in millimeters or inches). The electronic industry requires high quality large-area wafers that can be used for high-throughput device fabrication.

There are many methods, which are currently used individually or in combination for counting the number of atomic planes in graphene samples and for accessing the quality of graphene (e.g. presence of lattice defects or impurities (23, 24). Examples of these methods include micro-Raman spectroscopy (25, 26), optical microscopy (3-5), low-energy electron microscopy (LEEM) (27, 28), low-energy electron diffraction (LEED) (19, 27), atomic force microscopy (AFM) (1), scanning electron microscopy (SEM) (1), transmission electron microscopy (TEM) (1), scanning tunneling microscopy (STM) (29), photoelectron microscopy (PES) (27), angle-resolved photoelectron spectroscopy (ARPES) (27), photoemission electron microscope (PEEM) (27), Image J data analysis software (30), and reflection high-energy electron diffraction (RHEED) (31). At the same time, the existing methods have major limitations related to their slow, expensive, and non-automated measurement procedures. These methods can only be used for identification and quality analysis of small regions of SLG and sometimes FLG samples, but only on a few-micrometer range scale. Such methods become inadequate for scanning large-area graphene wafers on millimeter scale, which is required for industrial processes. Moreover, most of these techniques provide only rough estimates of the number of atomic layers.

Micro-Raman spectroscopy, the only non-destructive reliable and accurate technique, is limited to a small scanning spot size of a few micrometers. It would be time consuming to acquire Raman spectrum from the large area. The Raman spectroscopy method is limited to FLG films with the number of atomic planes smaller than n=5-7. For thicker films the Raman spectrum becomes too close to that from bulk graphite. There are also indications that Raman spectroscopy becomes less efficient for CVD graphene rather than for mechanically exfoliated graphene. The latter is explained by the increased $\pi - \pi$ bond stacking order (17), which intensifies 2D peak from the out-of-plane modes and preserves G peak of the in-plane modes, thus changing the well known G peak to 2D peak intensity ratio (32-34) and complicating interpretation of the Raman spectrum.

C. M. Nolen, G. Denina, D. Teweldebrhan, B. Bhanu and A.A. Balandin, University of California – Riverside, 2010

## I.  LARGE-SCALE GRAPHENE IDENTIFICATION

Here, we describe a process for the large-area graphene identification and quality control that is automated, cheap, robust, high-throughput and highly efficient. The technique is based on a combination of the modified optical contrast method with several optical filters and image processing algorithms. The calibration of the process is carried out using micro-Raman spectroscopy. The overall approach is illustrated in Figure 1. It shows a schematic of the process starting from a captured image via an optical microscope, micro-Raman calibration and image processing algorithm, which completes the recognition process. Below we provide a detailed step-by-step description of the process, which allows one to count the number of atomic planes (graphene layers) for a large-area and clearly identifies the borderlines between regions with different thickness (i.e. number of atomic planes).

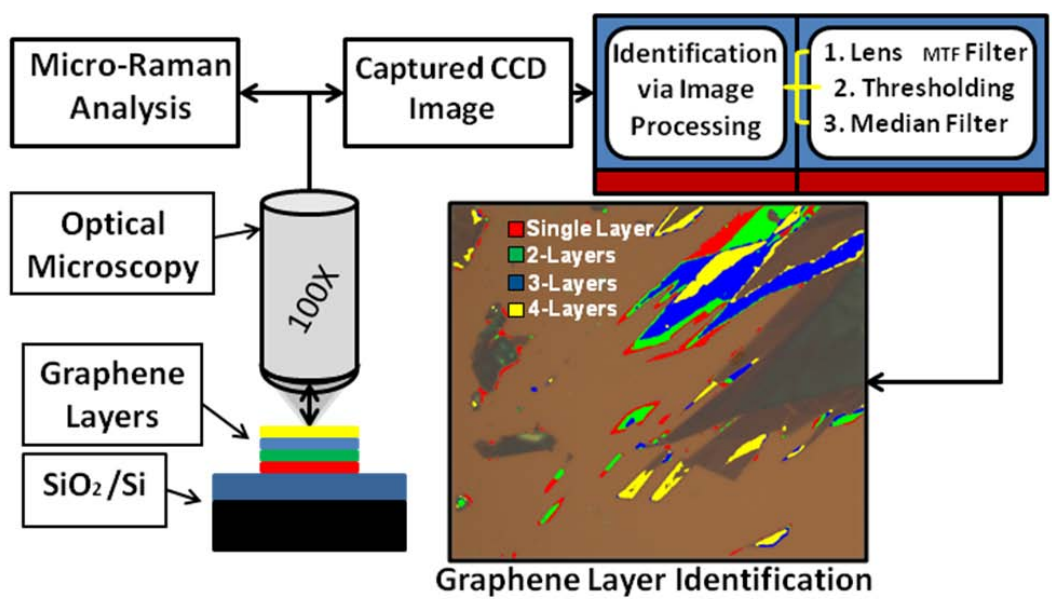

**Figure 1**: Schematic of the process for determining the number of layers in graphene and few-layer graphene films. The image of graphene sample is acquired via optical microscope, followed by micro-Raman calibration, background subtraction, light non-uniformity correction, and application of the original image processing algorithm to identify the regions with different number of atomic planes and map them with pseudo-colors.

To test the recognition process we prepared a large number of samples with graphene films. Graphene and FLG were produced by the standard mechanical exfoliation from HOPG and placed on top of the $SiO_2$ (300 nm)/ Si substrate (35, 36). It is known that the 300-nm thickness of $SiO_2$ allows one to visualize FLG regions under regular light conditions (~~12~~ 3-5). The high resolution optical microscopy images of the samples were captured by a digital camera attached to an optical microscope (Nikon Eclipse LV150) in "white" light produced by a quartz tungsten halogen light source.

***Step 1:*** We start by capturing two optical images. The first is of the substrate material (usually SiO2/Si but other substrates are also suitable) while the second image is of the FLG sample on the same substrate or different substrate made from the same material (see Figure 2). We intentionally selected samples with FLG regions containing different number of atomic plane n and having irregular shape boundaries. For convenience, we use the following notation: substrate without graphene – Image O and substrate with graphene – Image I.

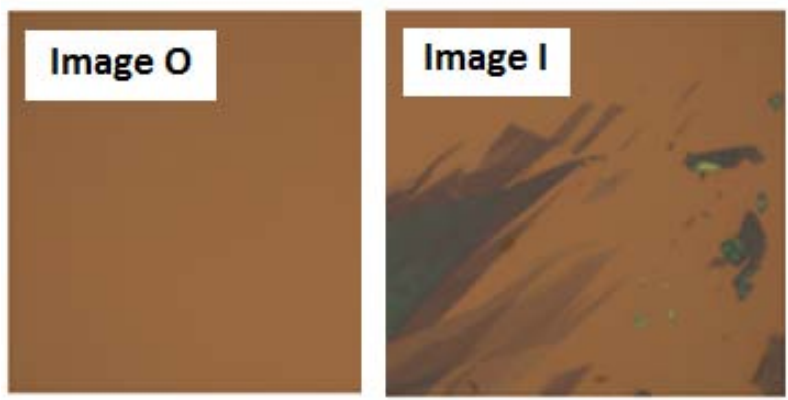

**Figure 2**: Microscopy images of the substrate (left panel) and FLG on substrate (right panel) obtained with 100X-objective. The image O is used for the background subtraction – a required step for the large area graphene identification. Image I is used for the overall graphene layer identification experiment.

***Step 2:*** We perform the calibration procedure, which can be done locally on a selected region where visual inspection suggest a presence of SLG, BLG, etc. This step involves collection of Raman spectrum from a few spots or performing a Raman line-scan. The Raman spectroscopy has proven to be very reliable for identification of FLG with $n$=1, 2, 3, 4 and 5 via deconvolution of 2D band and measuring the ratio of the intensities I(G)/I(2D). In most of cases, a single line scan is sufficient to identify at least one spot for each *n*. The coordinates of the spots, corresponding to *n*=1, 2, 3, 4, and 5 are recorded and correlated with the color information obtained in the previous step (both on Image O and Image I). This procedure accomplishes the labeling of several spots with the number of atomic planes (see Figure 3). This calibration step does not take much time because it is done locally and does not need to be repeated through the whole substrate coated with graphene or the whole wafer with CVD grown graphene. Moreover, once it is done for graphene on a certain substrate it can be omitted for other graphene samples on the same type of the substrates under the same illumination. We verify the Raman calibration via atomic force microscopy (AFM) inspection for randomly selected samples.

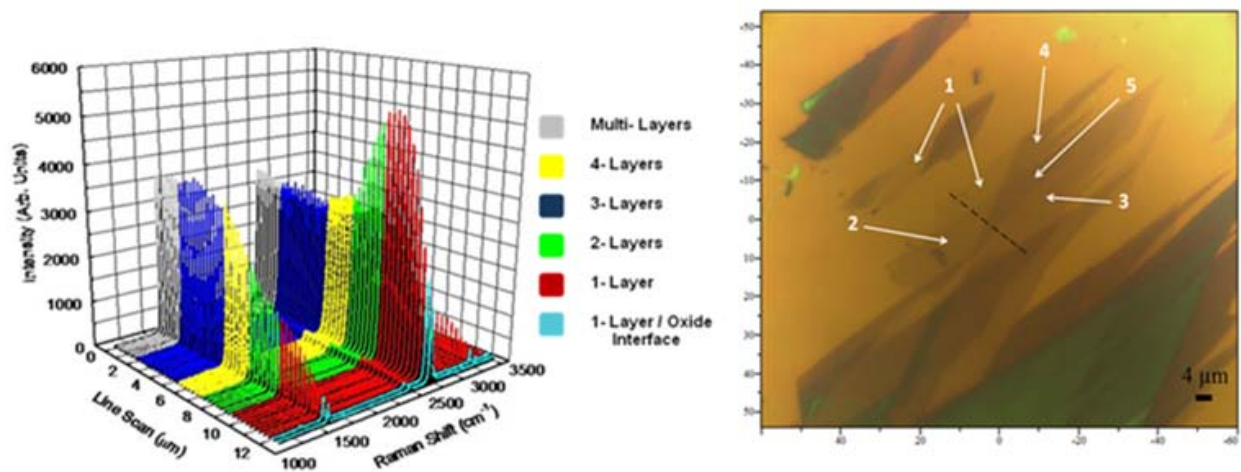

**Figure 3**: Raman line-scans showing the characteristic G peak and 2D band used for identification of the number of atomic plans n in FLG regions (left panel). The colors denote FLG with the number of planes varying from 1 to 4 and above. The Raman scans were taken along dotted line (12.5 μm length) indicated on the right panel. The white numbers label the number of atomic planes in different regions of FLG sample (right panel).

This procedure completed the preliminary (i.e. calibration steps) that do not have to be repeated for each new sample if the substrate and light conditions are kept the same. The following steps constitute the image processing algorithm applied to the captured optical images (Image I and Image O). In order to describe these steps ones needs to define and introduce a few concepts and variables, commonly used in image processing. We begin by noting that each optical image can be broken into a matrix of pixels with dimensions $M \times N$, where pixel row and column locations are in the range of $x, y \in 0 \le x \le M$, $0 \le y \le N$. Each pixel is assigned a light intensity in the range $I_{\min} \le I(x, y) \le I_{\max}$ for a given light source intensity. Here, $I_{\max}$ is the maximum intensity allowable (conventionally assumed to be 255), and $I_{\min}$ is the minimum intensity allowable (conventionally assumed to be 0), while $x$ and $y$ indicate the row and column (or coordinates) of the locations being computed. The intensity of each pixel can be represented as a combination of red (R), green (G), and blue (B) intensity values: $I(x, y) = [I_R(x, y), I_G(x, y), I_B(x, y)]$, where $I_R$ is the red intensity value, $I_G$ is the green intensity value and $I_B$ is the blue intensity value. With this in mind we can proceed to the next step.

**Step 3:** Since the main motivation for this research is development of the automatic scalable technique for the large-area graphene wafers, one needs to take into account the non-uniformity in wafer illumination. The optical images are taken using optical microscopes and unavoidably affected by the objective lenses, which do not produce uniform intensity of lighting throughout the image. The light is at its maximum intensity at the focal center and is the dimmest at the corner edges of the image. The illumination of the lighting non-uniformity is accomplished with the help of the reference substrate image (Image O). The intensity profile is found from Image O (see Figure 4) and then subtracted from Image I. This equalizes the lighting conditions over the whole substrate for the following image processing steps. The details of the non-uniform lighting removal process are given in the Methods section (see Section entitled *Non-Uniform Lighting Elimination and Supplemental Information*).

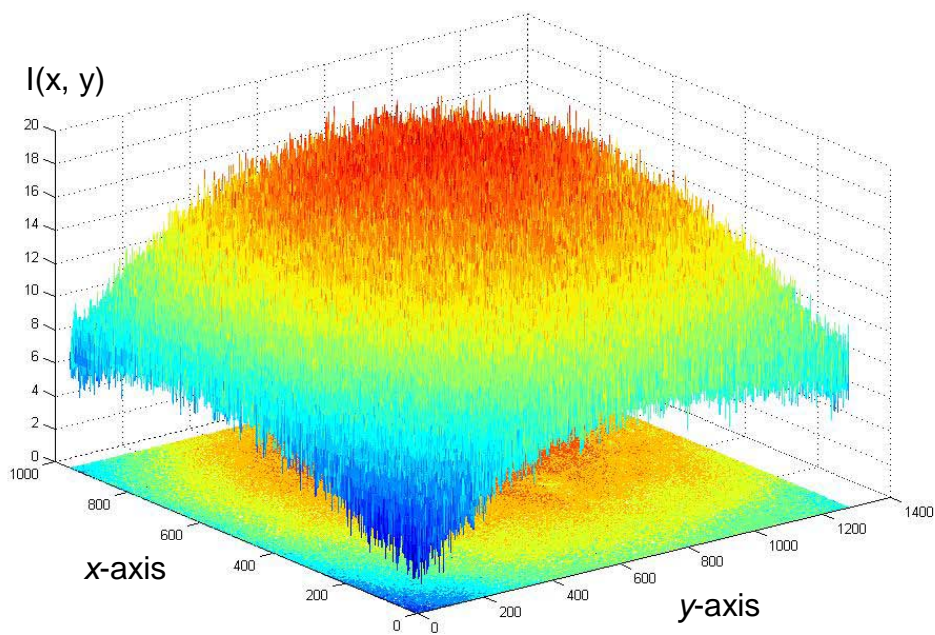

**Figure 4**: Non-uniform illumination intensity I(x, y), expressed in arbitrary units, is shown as a function of the substrate coordinates x and y. The non-uniformity caused by the circular confocal lens aberration is corrected by the lens modulation transfer function ($L_{MTF}$) filter introduced to the recognition procedure.

***Step 4:*** Once the uniformity of light illumination is achieved for the whole image one can extract the contrast information for different FLG regions (referenced to the background). To conduct such a process for Image I, we define the red, green, and blue (RGB) values for each pixel of the image. From the step 2, we know what RGB values correspond to regions with $n$=1, 2, 3, 4 or 5 (see Figure 5). This determines the range of RGB values that ensure that the region has the number of atomic planes within $n$=1-5. Using this information we identify regions of FLG throughout the whole image or wafer. After we have specified all FLG regions of interest (e.g. with $n$<4) in pixels within $M \times N$ we can exclude all other thicker regions (e.g. with n>4). The exclusion is based on subtracting all points of the image that have RGB levels above or below the allowable RGB previously specified for each $n$ (see Figure 5). The detail algorithm for the RGB assignment and image processing exclusion of regions that do not belong to the needed n range are described in the Methods section (see *Background Subtraction*).

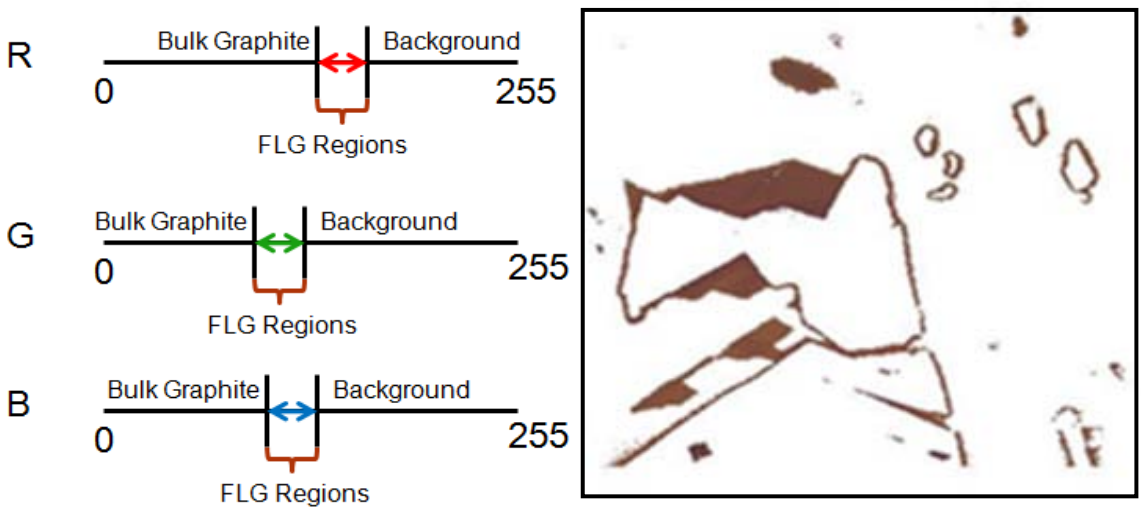

**Figure 5**: Range of the red, green and blue (RGB) light intensity values that corresponds to the FLG regions (left panel). The RGB ranges can be used for the background subtraction and exclusion of the regions with thicker films (e.g. bulk graphite). Optical image (100X magnification) after RGB processing, which restricted the light intensity values to FLG regions (right panel). Only the dark regions consist of FLG with $n$=1-4, the rest of the regions (white) are either substrate or thicker films (e.g. bulk graphite).

***Step 5:*** We can now refine the recognition process and perform identification of each graphene layer (with specific $n$) from the determined FLG regions. To accomplish this task we start by converting the RGB data (defined for FLG regions in Image I) that contains 3 values per pixel to the grayscale that contains 1 value per pixel. The latter is accomplished through the process called segmentation (see Methods section entitled *Graphene Layer Identification*). We label the grayscale contrast range for FLG regions, defined as $\Sigma \Delta I_n$, and find the minimum – maximum boundary range for the graphene layers with specified $n$ (see Figure 6). The intensity range for graphene layers with a given $n$ is labeled as $\Delta I_n$ (it is contained within $\Sigma \Delta I_n$ range). The use of grayscale can only be efficient for FLG regions after removal of the background and regions that correspond to the thick graphitic films.

The optical adsorption of each graphene layer for different brightness intensities is shown in Figure 6, where $\Delta I_n$ contains the range of the light intensity values associated with a specific graphene layer of interest (specified by a given $n$) and $\Sigma \Delta I_n$ shows the light intensity range of values for the entire FLG region. The range of these light intensity values depends on the brightness of the light source of the optical microscope. The dependence of $\Delta I_n$ on the intensity of the light source is important for this automated identification process.

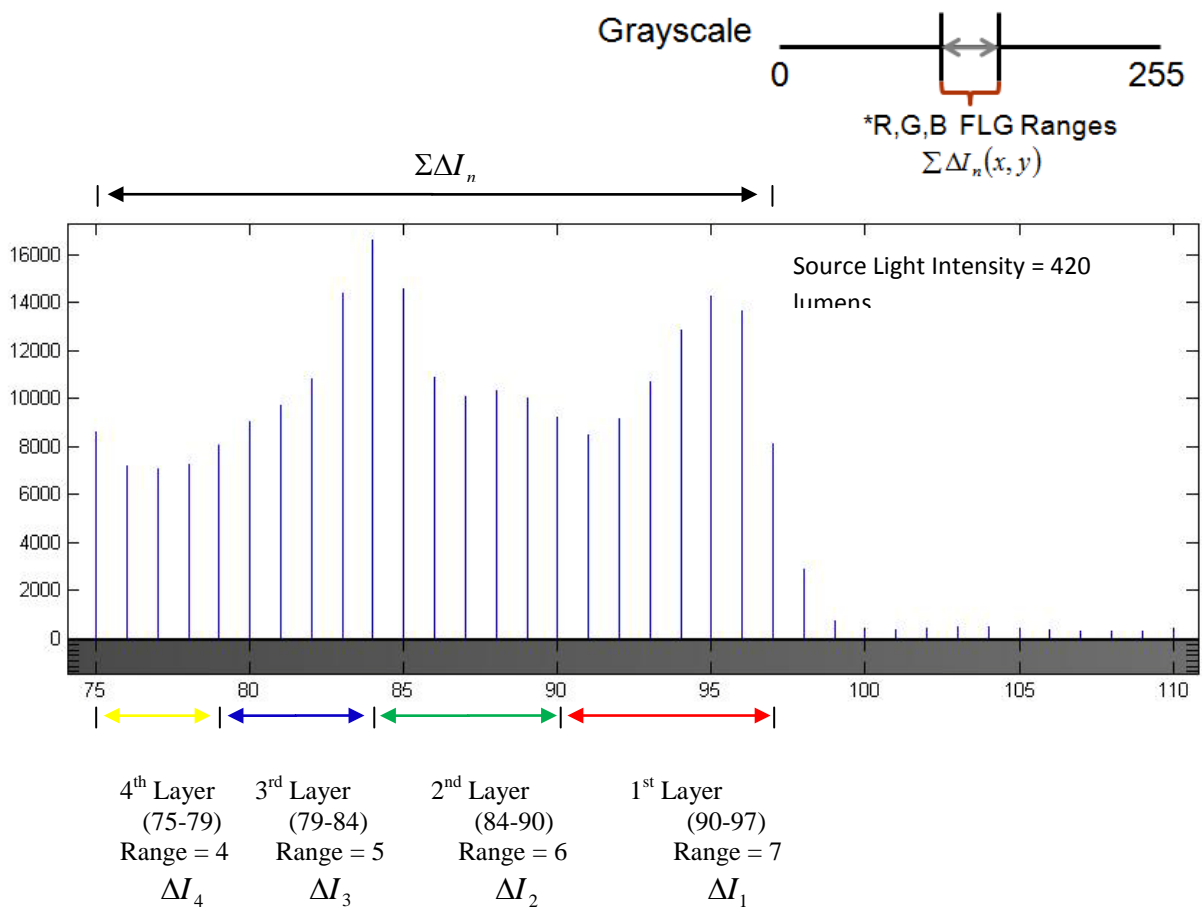

**Figure 6**: Illustration of the conversion process of RGB values to the grayscale (upper panel). The grayscale allows for easier identification of the regions with a specific number of atomic planes $n$. The grayscale range of FLG is sectioned into the respective light intensity ranges for each individual layer of graphene with a given $n$ (lower panel).

For clear visual recognition, the unique pseudo colors are assigned to the contrast ranges $\Delta I_n$ for each graphene layer with a given $n$. This is done by further filtering out separately the regions for SLG, BLG, and FLG with $n=3$ and 4 from the grayscale FLG region. In Figure 7, we show how such filtering results in the separated regions. The white spots correspond to the regions with a specific number of atomic planes *n*. The entire M x N transparent image with identified pseudo colored regions is then laid on top of the original optical image (Image I), for visual identification of FLG regions with desired *n*. The mathematical details of the process are described in the Methods section (see *Graphene Layer Identification*).

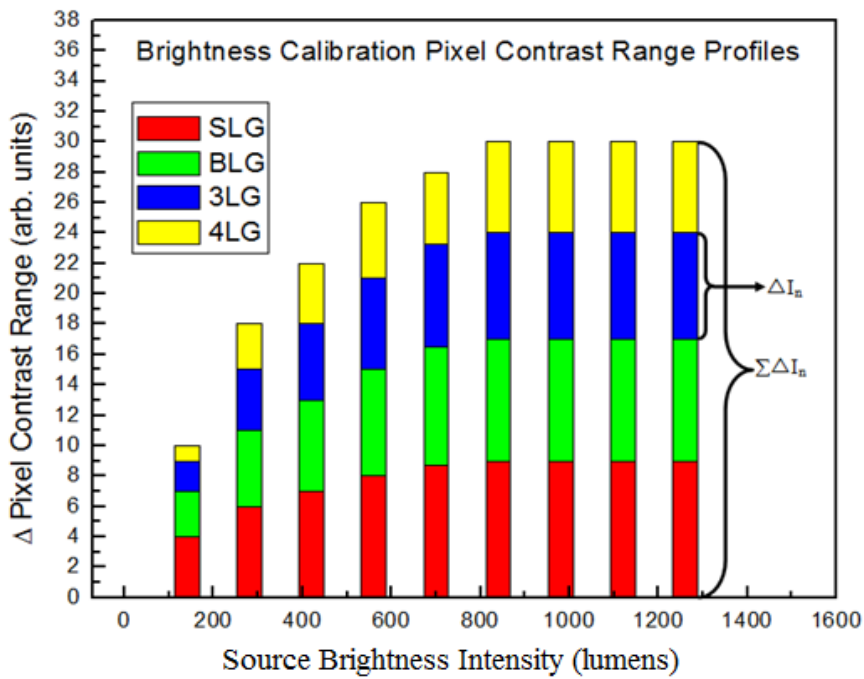

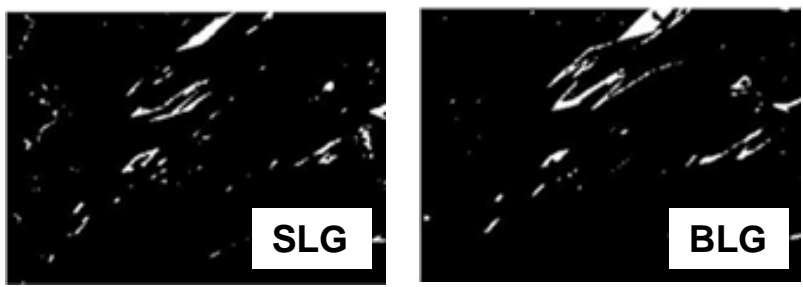

**Figure 7**: Optical brightness ranges $\Delta I_n$ associated with a specific graphene layer of interest (defined by *n*) and combined range $\Sigma \Delta I_n$ for the entire FLG region shown as functions of the source brightness (upper panel). The white spots on the sample surface are identified regions of SLG (n=1) and BLG (n=2) (lower panel). The dark background is the rest of the sample surface (i.e. regions of substrate without graphene or regions with other thicknesses). The presented technique can be used for wafer size samples without any major modification or processing time increase.

**Step 6:** The graphene identification procedure is completed with an application of the median filter and utilization of pseudo colors for better visualization. The median filtering step involves the statistical pixel-to-pixel neighborhood analysis to improve the image appearance and accuracy within the identified region and clarify the boundaries between any two regions with different number of atomic planes $n$ (see also the *Median Filter* in the Methods section). The median filter allows one to remove the high frequency impulse noise commonly known in image processing as "salt and pepper" noise. In our approach, this noise may cause the identified regions of graphene to appear patchy reducing the accuracy when determining the borders of the regions. After the filtering process, we assign the pseudo-color to each region with a given $n$ *value*, and present the final result on the sample map (see Figure 8). This map clearly marks the number of atomic planes at each location of the sample surface by color: red, green, blue or yellow. The remaining brown regions are the substrate itself without graphene flakes while the dark regions are the thicker graphite films.

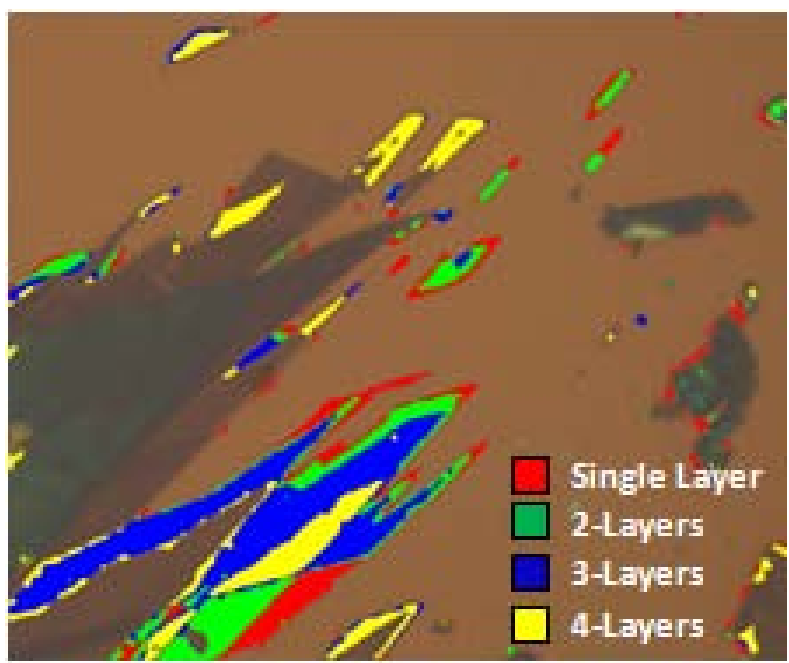

**Figure 8:** Statistical nanometrology layer counting analysis is performed by sectioning out a specific individual graphene layer different from the bulk, substrate, and other graphene layers.

C. M. Nolen, G. Denina, D. Teweldebrhan, B. Bhanu and A.A. Balandin, University of California – Riverside, 2010

It is easy to see that the approach can be extended to the wafer size or rolls of CVD graphene on flexible substrates. Since the only size limitation is the area of the optical image our approach is suitable for industry scale high-throughput applications. The high speed of the image processing algorithms allows for the *in situ* identification of the number of atomic planes. As a result, the throughput for the industrial scale inspection of many wafers will be determined by the speed of mechanical motion of the wafers to and from the light source. A similar scale of the graphene identification cannot be achieved with Raman spectroscopy. The two-dimensional Raman scan of the whole sample surface will take an extremely long time (the spectroscopic data accumulation for each point on the sample surface takes from ~1 minute to ~30 minutes with conventional spectrometers). The lateral resolution of Raman spectrometers is determined by the laser spot size, which is on the order of 0.5 – 1.0 μm.  In our approach, the Raman spectroscopy is used only for the calibration of the process and can be done locally on a few stops or via a line scan.

The versatility of our metrology technique opens a door for a plethora of experimental and industrial applications. It can be applied to a number of various substrates and graphene samples produced by different methods (37-39). Instead of micro-Raman spectroscopy, it can use other calibration techniques. We have tested the method on a large number of graphene samples produced by mechanical exfoliation. In some cases we intentionally used contaminated substrates and FLG flakes that had a large thickness variation (from SLG to bulk graphite). Our technique worked fine for all examined substrates. Moreover, we have tested our approach for another type of the atomically thin materials – topological insulators of the bismuth telluride family (40-45). The "graphene-like" exfoliated atomically thin films of $Bi_2Te_3$ and $Bi_2Se_3$ (40-43) were placed on top of $Si/SiO_2$ substrates. The graphene identification technique performed for this type of samples as well. The accuracy of our technique can be enhanced further by application of other image enhancing, error reducing algorithms implemented in different software packages. The image processing applications in semiconductor industry have already helped to achieve major improvements in materials processing and chip fabrication at reduced cost (46). The proposed large-scale graphene identification and quality control technology can become particularly useful for the newly developed graphene synthesis techniques (47, 48) and graphene practical applications in heat spreaders, interconnects and analog electronics (49-52).

## II. CONCLUSIONS

We described a fast and fully automated approach for determining the number of atomic planes (i.e. number of layers) in graphene samples. The procedure allows for *in situ* identification of the borders of the regions with the same number of atomic planes. It is based on an original image processing algorithm, which utilizes micro-Raman calibration (on a local area), light background subtraction, lighting non-uniformity correction, and the color and grayscale image information processing for each pixel. Our approach works for various substrates, and can be applied to the mechanically exfoliated, chemically derived, deposited or epitaxial graphene on an industrial scale. The automatic identification procedure is described step-by-step with the mathematical details summarized in the methods sections.

### *Acknowledgements*

This work was supported by the ONR project on the Graphene "Quilts" for Thermal Management of GaN Power Electronics. A.A.B. also acknowledges partial support from SRC - DARPA Focus Center Research Program (FCRP) through its Center on Functional Engineered Nano Architectonics (FENA) and from DARPA Defense Microelectronics Activity (DMEA) under agreement number H94003-10-2-1003.

## III. METHODS

This section is divided into the sub-sections, which provide details for specific steps of the large-scale graphene identification and quality control technique.

### III.A. Non-Uniform Lighting Elimination

This description is pertinent to Step 3 of the procedure. To eliminate the non-uniform lighting across Image I, we apply a special filter, which subtracts the light intensity extracted from the background Image O. This is based upon the assumption that under perfect condition, the color intensity of the substrate is uniformly the same across the image. The whole procedure is presented step by step in Figure 9. It starts with accumulation of the typical light intensity distribution for Image O along the x or y axis. This intensity distribution is non-uniform with the maximum attained usually around the center of the image. The distribution is modified by subtraction of the uniform background. The resulting non-uniform part is inverted and stored for further use with Image I. The next step is accumulation of the typical intensity distribution for Image I (the actual graphene sample on the substrate). The addition of the inverted light intensity, obtained for the reference Image O, to the intensity distribution in Image I results in the corrected intensity distribution for Image I with eliminated lighting non-uniformity (shown in the lower right panel in Figure 9).

Mathematically, this process is described as an application of the lens modulation transfer function ($L_{MTF}$) filter (53). The filter corrects the circular lens aberration produced by the Gaussian-like distribution of non-uniform light intensity in both the x and y planes of Image I (see Figure 4 and 9). The application of the $L_{MTF}$ filter is performed with the equation

$$I_{n,C\in R,G,B}(x,y) = I_{C\in R,G,B}(x,y) - L_{MTF} \qquad (1)$$

for each value $I_R$, $I_G$, $I_B$ where $L_{MTF} = O_{C\in R,G,B}(x,y) - \min(O_{C\in R,G,B})$. The intensity function $I_n$ now contains the corrected image with the evenly distributed light intensity across the entire image.

C. M. Nolen, G. Denina, D. Teweldebrhan, B. Bhanu and A.A. Balandin, University of California – Riverside, 2010

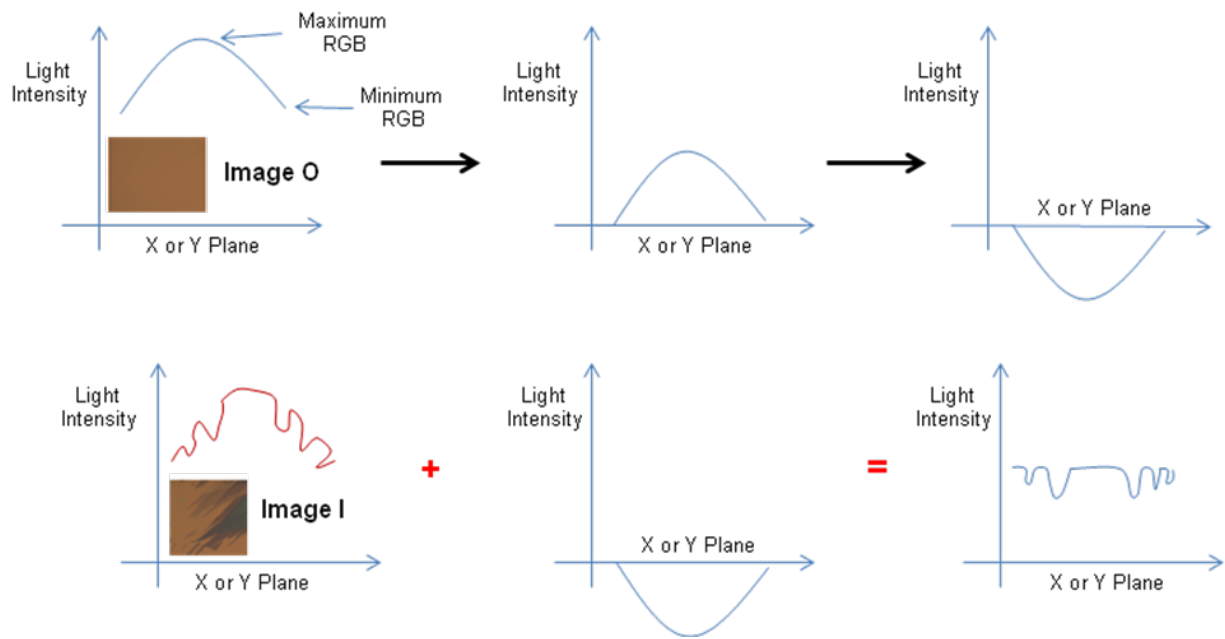

**Figure 9:** The upper left panel shows a typical non-uniform light intensity distribution for Image O along the *x* or *y* axis. The upper central panel is the non-uniform light intensity, which is left after subtraction of the uniform background lighting. The upper right panel shows the inverted light intensity obtained in the previous step. This completes the processing of data extracted from Image O. The lower left panel is a typical intensity distribution for Image I along the *x* or *y* axis. The modulation of the intensity profile is due to the presence of graphene flakes with different number of atomic planes. The addition of the inverted light intensity, obtained for Image O (upper right panel), to the Image I light intensity results in the final intensity distribution for Image I with eliminated lighting non-uniformity. The final result is shown in the lower right panel.

### III.B: Background Subtraction

This procedure is pertinent to Step 4. In order to subtract the background, we subtract the RGB value from all pixels that correspond to the same location in Image O and Image I. If the result is ~ 0, then the pixel in Image I is assumed to be a background pixel, in which case we change that RGB pixel value to white corresponding to (0, 0, 0). If the result is 1, then the pixel in Image I is assumed to be not a background pixel, in which case we do not change that RGB pixel value, and retain its original RGB value. This is accomplished by performing the following procedure

$$M(x,y) = \begin{cases} 0 \ \ if \ \ O_{C \in R,G,B}(x,y) - I_{C \in R,G,B}(x,y) \approx 0 \\ 1 \ \ if \ \ O_{C \in R,G,B}(x,y) - I_{C \in R,G,B}(x,y) \neq 0 \end{cases}, \qquad (2)$$

where M contains the filter resulting from Image I with the substrate background subtracted. Next, by using the light contrast information from Figure 2, we can restrict each RGB value to only allow the light intensity range for FLG regions shown in Figure 5.

### III.C: Graphene Layer Identification

This procedure is pertinent to Step 5. To perform the identification of graphene layers (i.e. distinguish SLG from BLG, etc.), each pixel in the entire image needs to be "segmented" (54) from RGB colors into a grayscale color. This makes the graphene layer extraction process simpler by converting three RGB values per pixel into one grayscale value per pixel, which comprises of a different percentage of each RGB value. The grayscale conversion is completed by changing the M regions of interest to grayscale and changing all other pixels that are not within the FLG contrast range to white by performing the following actions

$$I_{n,Gry} = 0.30I_{n,R} + 0.59I_{n,G} + 0.11I_{n,B}$$

$$I_n(x,y) = \begin{cases} 255 & if \ M(x,y) = 0 \\ I_{n,Gry}(x,y) & if \ M(x,y) = 1 \end{cases}. \qquad (3)$$

Here $I_n$ is an image containing only FLG. Next, the regions with specific number of atomic planes n (e.g. SLG or BLG) are determined from their grayscale light intensity range acquired from the optical Image I. The latter is achieved by applying the neighborhood thresholding (54), which allows one to extract the light intensity contrast range for each individual graphene layer with given $n$ by performing the following operations

$$\Sigma \Delta I_n(x,y) = \begin{cases} 1 & L1_{min} \leq \Delta I_1(x,y) \leq L1_{max} \\ 2 & L2_{min} \leq \Delta I_2(x,y) \leq L2_{max} \\ 3 & L3_{min} \leq \Delta I_3(x,y) \leq L3_{max} \\ 4 & L4_{min} \leq \Delta I_4(x,y) \leq L4_{max} \\ 0 & other \end{cases}. \qquad (4)$$

Here $\Sigma \Delta I_n$ represents the summation of the light intensity ranges for FLG regions containing each of the grouped layers and $\Delta I_n$ is the light intensity range for a specific graphene layer of interest, L1 is SLG, L2 is BLG, and L3, L4 are FLG with $n=3$ and $n=4$, respectively. The minimum and maximum values span the light intensity threshold range for each graphene layer (with given $n$). This process is repeated over the entire range of the source light intensities in order to provide a calibration lookup table for the image processing algorithms. As a result, the image processing algorithm can section out the graphene light intensity ranges for any light source intensity. The unique pseudo colors are then assigned to each graphene layer and overlaid atop of the original image for clear visual identification. The unique colors are assigned to the rest of the image, which include regions of the substrate or bulk graphite.

### III.D: Median Filter

This procedure is pertinent to Step 6. To achieve the precision sufficient for the large-scale industrial implementation and automation, a median filter (54) is applied to eradicate the high frequency impulse noise. The median filter for each individual layer is implemented with the help of formula

$$M_F = \{ I_{T_{n_{jk}}} \mid j \in \{1,2,...,W\} \ and \ k \in \{1,2,...,H\} \} , \tag{5}$$

where $M_F$ is a median filter of size $W \times H$ for a neighborhood of pixels centered at $I_{T_n}(x,y)$. The median element of the window $M_F$ is given by

$$I_{F_n}(x,y) = \begin{cases} M_{F_{SORT}}\left[ \dfrac{m}{2} \right] & for \ an \ even \ n \\ M_{F_{SORT}}\left[ \dfrac{m}{2} + 1 \right] & for \ an \ odd \ n \end{cases} , \tag{6}$$

where $M_{F_{SORT}}[i]$, $i = 1$, $m$, $m = W \times H$ and $I_{F_n}$ is the resulting graphene layer of interest (with given $n$) after the impulse noise is removed. The median filter analyzes a set number of pixels in a user-defined matrix region to find the median value of the region currently being inspected. After the operation is performed, the filter is shifted to the next user-defined matrix region until the entire image is analyzed. We tuned our matrix size to effectively eliminate noise while maintaining the high accuracy for the identified graphene regions (with given $n$) on the order of a few nanometers.